\documentclass[11pt]{article}

\usepackage[final]{acl}

\usepackage{times}
\usepackage{latexsym}

\usepackage[T1]{fontenc}

\usepackage[utf8]{inputenc}

\usepackage{microtype}

\usepackage{inconsolata}

\usepackage{graphicx}

\usepackage{balance}  
\usepackage{array}
\usepackage{booktabs}
\usepackage{multirow}
\usepackage{multicol}
\usepackage{makecell}
\usepackage{tabularx}
\usepackage{colortbl}
\usepackage{subcaption}
\usepackage[breakable]{tcolorbox}
\usepackage[dvipsnames]{xcolor}
\usepackage{xurl}
\usepackage{enumitem}
\usepackage[table]{xcolor}
\usepackage{authblk}
\usepackage{makecell}

\usepackage{fvextra} 

\newcommand{\sysname}{MemeBuddy}

\definecolor{tableline-gray}{gray}{0.7}
\definecolor{light-gray}{gray}{0.9}
\definecolor{brandblue}{RGB}{87, 178, 255}

\definecolor{narrator-color}{RGB}{87, 178, 255}
\definecolor{femalevoice-color}{RGB}{255, 217, 231}
\definecolor{malevoice-color}{RGB}{199, 231, 255}

\definecolor{tcolorbox-background}{gray}{0.98}
\definecolor{tcolorbox-frame}{gray}{0.4}

\title{MemeBuddy: Dialog-Style Audio Representations for Engaging Non-Visual Meme Experiences}

\author{
  \textbf{Chirag Bhansali\textsuperscript{1}},
  \textbf{Vikas Ashok\textsuperscript{2}},
  \textbf{Hae-Na Lee\textsuperscript{1}}
  \\
  
  \begin{tabular}[t]{c c}
    \textsuperscript{1}Department of Computer Science and Engineering & \textsuperscript{2}Department of Computer Science \\
    Michigan State University & Old Dominion University \\
    East Lansing, MI, USA & Norfolk, VA, USA \\
    \texttt{\{bhansal4, leehaena\}@msu.edu} & \texttt{vganjigu@odu.edu} \\
  \end{tabular}
}

\begin{document}
\maketitle

\begin{abstract}
Image memes are a pervasive form of online communication, widely used to convey humor, opinions, and cultural references. Prior work has explored making memes accessible to blind users, primarily through auto-generated descriptive captions. While these approaches improve comprehensibility and sometimes incorporate prosodic or emotional cues, they often fail to capture the humor, narrative structure, and contextual nuances that make memes engaging. We present \sysname{}, a system that models memes as dialog, generating structured, multi-turn audio representations using role-based speakers. \sysname{} reinterprets a meme as a conversation between two speakers, integrating extracted meme text with contextual knowledge implicitly inferred by a multimodal LLM (e.g., recognition of common meme templates and cultural references) to convey intent, timing, and implicit meaning through conversational interaction. We evaluate \sysname{} in a user study with 14 blind participants. Results show that dialog-style meme representations consistently improve engagement and user satisfaction compared to caption-style descriptions, while maintaining comparable comprehension.          
\end{abstract}

\section{Introduction}

Image \textit{memes} are humorous and culturally referential visual artifacts that have become a dominant form of online communication across social media, forums, and messaging platforms~\cite{davison2012language, molina2020makes, bauckhage2011insights, kostadinovska2018internet, morina2022webscale}. Users rely on memes to express opinions, share experiences, and participate in collective cultural discourse~\cite{chen2012creation, blommaert2015conviviality}. Ensuring equitable access to memes is therefore critical, particularly for blind users who interact with content through synthesized speech via screen readers (e.g., JAWS, VoiceOver, NVDA).

However, meme accessibility remains limited. Screen readers primarily depend on \textit{alt text}, which is often missing, incomplete, or insufficient for conveying meaning of images to the blind users~\cite{voykinska2016how, gleason2019its, webaim2026million}. To compensate, blind users increasingly rely on AI-based tools (e.g., ChatGPT, Seeing AI, Be My AI) that generate image descriptions on demand~\cite{sharma2025before, adnin2024look}. While these systems improve \emph{comprehensibility}, they typically produce static, caption-style outputs that fail to capture the humor, timing, and contextual nuance central to memes~\cite{dynel2016has, vasquez2021cats}. As a result, the users often receive functionally correct but experientially diminished representations.

Recent work has explored enhancing engagement through emotional narration~\cite{chen2025memespeak}. However, the approach largely retains a single-speaker, descriptive paradigm. We argue that the limitation lies not only in \emph{what} is conveyed, but in \emph{how} it is structured. Memes are inherently social and interpretative; their meaning often unfolds through reactions, timing, and shared understanding. This motivates modeling memes as \emph{dialogs} rather than static descriptions.

\begin{table*}[h!]
\centering
{\small
{\def\arraystretch{1.1}
\begin{tabular}{ m{2.7cm} >{\raggedright\arraybackslash}m{4.3cm} >{\raggedright\arraybackslash}m{4.0cm} >{\raggedright\arraybackslash}m{3.4cm}}
\toprule
 \multirow{3}{*}{\textbf{Meme Image}} & \textbf{Baseline (BL)} & \textbf{Commentator--Commentator (CO)} & \textbf{Commentator--Character (CH)}\\ \arrayrulecolor{tableline-gray}\cline{2-4}
 & One speaker & Two speakers & One speaker and \newline a meme character \\
 \arrayrulecolor{black}
 \midrule
 \includegraphics[scale=0.42]{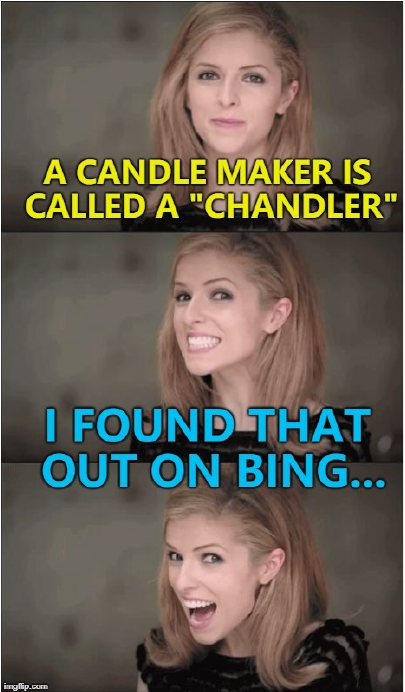}
 & \colorbox{femalevoice-color}{\textbf{Narrator:}} Alright, I've got a fun three-panel meme here featuring the actress Anna Kendrick. 
In the top panel, she's looking right at the camera with a casual, kind of amused expression. The text over her in bright yellow says, `A CANDLE MAKER IS CALLED A "CHANDLER"'. 
Then in the middle panel, she leans in a bit with this mischievous, excited smile, baring her teeth just a little. The text here changes to blue and reads, `I FOUND THAT OUT ON BING...'.
Finally, in the bottom panel, there's no text at all. It's just Anna bursting into a big, playful laugh, looking completely proud of herself.
So, the joke is basically a giant pun. If you put the keywords from the top and middle panels together - `Chandler' and `Bing' - you get Chandler Bing, the famous character from the TV show Friends. Her totally cracking up at her own dad joke is what really sells it!
 & \colorbox{malevoice-color}{\textbf{Host1:}} This is a three-panel meme featuring actress Anna Kendrick, setting up a clever little pun.
 
\colorbox{femalevoice-color}{\textbf{Host2:}} She's got great facial expressions. What's the setup?

\colorbox{malevoice-color}{\textbf{Host1:}} In the first panel, she has a pleasant, normal expression with text that says, `A candle maker is called a Chandler.'

\colorbox{femalevoice-color}{\textbf{Host2:}} Alright, fun little trivia fact. What's next?

\colorbox{malevoice-color}{\textbf{Host1:}} The second panel shows her making a cheesy, awkward grin. The text adds, `I found that out on Bing...'

\colorbox{femalevoice-color}{\textbf{Host2:}} Wait... Chandler... Bing. Oh, wow.

\colorbox{malevoice-color}{\textbf{Host1:}} Exactly. And the final panel is just a picture of her laughing enthusiastically at her own joke.

\colorbox{femalevoice-color}{\textbf{Host2:}} So basically, it's just an elaborate dad joke hiding a `Friends' reference, and she is absolutely thrilled with her own genius.

& \colorbox{malevoice-color}{\textbf{Host:}} Today we're looking at a three-panel meme featuring a famous actress dropping some unexpected trivia.

\colorbox{femalevoice-color}{\textbf{Character:}} Hi, \textit{I'm Anna Kendrick!} In the top panel, I'm just looking perfectly normal and polite, sharing a fun fact: `A candle maker is called a Chandler'.

\colorbox{malevoice-color}{\textbf{Host:}} Then in the middle panel, you're making this hilariously forced, awkward smile with the text, `I found that out on Bing...'

\colorbox{femalevoice-color}{\textbf{Character:}} Exactly. And by the bottom panel, I'm just flashing a massive, exaggerated, cheesy grin with no text at all to let the joke land.

\colorbox{malevoice-color}{\textbf{Host:}} Chandler Bing. The ultimate `Friends' dad-joke, delivered with the perfect sarcastic smirk.\\
\arrayrulecolor{black}
\bottomrule
\end{tabular}
}
}
\vspace{-5pt}
\caption{Illustration of \sysname{}-generated audio representations for a given meme image.}
\label{table:meme_teaser}
\end{table*}

Towards this, we present \sysname{}, a system that generates dialog-style audio representations of memes using role-based speakers. Specifically, \sysname{} reimagines a meme as a structured, multi-turn conversation between two speakers, enabling the gradual unfolding of humor, intent, and cultural context. As illustrated in Table~\ref{table:meme_teaser}, \sysname{} supports two dialog variants: (i) commentator–commentator (CO), where two speakers collaboratively interpret a given meme, and (ii) commentator–character (CH), where a speaker interacts with a character from the meme. A caption-style baseline (BL) is also supported for comparison.

We evaluate \sysname{} through a user study with $14$ blind participants. Results show that dialog-style representations consistently improve engagement, immersion, and enjoyment compared to caption-style descriptions, while maintaining comparable comprehension. Our findings further reveal trade-offs between conversational coherence and expressiveness, offering insights into how non-visual content representations can be made both accessible and engaging for blind users.

In summary, we contribute: (i) Novel dialog-based audio representations using role-based speakers for engaging meme consumption; and (ii) An empirical evaluation demonstrating that dialog structure consistently improves meme engagement while preserving comprehension.

\section{Related Work}
Our work closely relates to prior efforts in making visual media accessible, as well as research on meme understanding and accessibility.

\subsection{Accessibility of Visual Web Content}

Prior work has explored improving blind and low vision users' access to visual web content, including images, memes, GIFs, and videos~\cite{gleason2019its,prakash2024improving,lee2021towards}. Early systems, such as \textit{WebInSight}, generated alt text using web context and OCR, highlighting both the prevalence of missing descriptions and the potential of automation~\cite{bigham2008webinsight}. Later approaches introduced automatic image descriptions (e.g., Automatic Alt-text~\cite{wu2017automatic}), yet studies show that both user-provided and AI-generated descriptions are often sparse or misaligned with users' needs~\cite{morris2016with, voykinska2016how, gleason2019its}. Large-scale audits confirm that many images still lack meaningful alt text~\cite{webaim2026million}.

Beyond basic descriptions, research has examined richer representations of visual content. Previous works highlight the need for flexible, context-sensitive descriptions and interactive access to visual structure~\cite{morris2018rich, stangl2021going, lee2021image}. Accessibility efforts have also extended to dynamic media, including memes, GIFs, and videos, through approaches such as emotion-aware captions, expressive narration, and hierarchical summaries~\cite{gleason2019making, prajwal2019towards, chen2025memespeak, daele2024making, zhang2022ga11y}. However, AI-based tools, while increasingly used in practice, still exhibit limitations in usability and experiential quality~\cite{penuela2024investigating, adnin2024look, sharma2025before}.

Overall, prior work primarily focuses on \emph{semantic access}, i.e., helping users understand what is depicted, while retaining caption-style paradigms. Less attention has been given to preserving \emph{social and experiential qualities} of content. Our work addresses this gap by modeling memes as dialog, enabling engagement as well as comprehension.

\subsection{Memes and their Accessibility}

Internet memes are a multimodal and culturally-grounded form of communication that combine visual and textual elements to convey humor, opinions, and social meaning~\cite{davison2012language, molina2020makes, bauckhage2011insights, blommaert2015conviviality, kostadinovska2018internet, morina2022webscale, chen2012creation}. Understanding memes requires reasoning over visual metaphor, inter-textual references (e.g., celebrities, popular culture), and contextual knowledge~\cite{vasquez2021cats, hwang2023memecap, joshi2024contextualizing}, making them more complex than standard images.
Prior works have addressed meme understanding through captioning, explanation, and multimodal reasoning~\cite{hwang2023memecap, sharma2023what, zhong2024multimodal}, as well as emotion and metaphor analysis~\cite{xu2022metmeme, sharma2024emotion}. These efforts highlight the inherently inferential nature of memes, where meaning extends beyond literal content.

Prior accessibility work has focused on enriching meme descriptions. Gleason et al.~\cite{gleason2019making} proposed structured descriptions beyond OCR text, while Prajwal et al.~\cite{prajwal2019towards} incorporated facial-emotion cues. More recently, \textit{MemeSpeak}~\cite{chen2025memespeak} demonstrated that emotional speech can enhance meme narration.
However, existing approaches largely remain within a \emph{descriptive paradigm}, typically relying on single-speaker narration. This overlooks the inherently social and temporal nature of memes, where meaning unfolds through timing, interaction, and interpretation. As a result, prior work improves \emph{understandability} but does not fully address \emph{engagement}. Our work addresses this gap by modeling memes as dialogs through \emph{structural re-representation}. \sysname{} generates multi-turn, role-based conversational representations that enable meaning to gradually unfold through interaction, providing a more engaging non-visual meme experience while complementing prior approaches in this area.

\section{\sysname{} Design}

The design of \sysname{} was informed by both prior research on dialog as an effective medium for human-information interaction~\cite{ainsworth2004multiple, chi2017why, ghazarian2021discol, mizukami2016analyzing, shuster2020image} and insights from an IRB-approved focus group study. The study involved seven participants with complementary expertise: two authors, a psycholinguist, a media researcher, a social computing researcher, and two blind university students with extensive experience in meme consumption and creation on social media platforms.

The focus group was conducted as a structured design elicitation session. The participants were guided using seed themes derived from prior work on accessible content representation and conversational interfaces, and were asked to collaboratively explore how visual memes could be translated into engaging non-visual audio formats. Discussions centered on key design dimensions, including number of speakers, distribution and timing of information across turns, mechanisms for conveying humor and sarcasm, and desirable prosodic characteristics of voices. All sessions were audio-recorded and transcribed for subsequent analysis.

We analyzed the transcripts using a qualitative coding process following open and axial coding procedures~\cite{saldana2011fundamentals}. In the open coding phase, recurring concepts (e.g., ``turn-taking improves clarity'', ``characters increase immersion but may confuse roles'') were identified. These codes were then grouped during axial coding into higher-level design principles. Two dominant dialog structures emerged: (i) a commentator--commentator format emphasizing clarity and conversational coherence, and (ii) a commentator--character format emphasizing immersion and expressiveness. Additionally, participants consistently requested a caption-style fallback to ensure reliability when dialog representations are insufficient. We operationalized these insights as prompt-level constraints in \sysname{} (Appendix~\ref{meme_representation_generation_prompts_sec}).

At a system level, \sysname{} uses controlled prompts to a multimodal LLM to transform an input meme image into structured outputs containing both dialog and corresponding text-to-speech (TTS) instructions. These outputs are rendered as audio using expressive TTS engines. The strategies are described next.

\subsection{Commentator-Commentator (CO) Design}

The group strongly preferred two-speaker dialogs over multi-speaker alternatives, citing reduced cognitive load, improved coherence, and avoidance of unnecessary complexity. We encoded these preferences directly into the LLM prompt used for CO dialog generation (Table~\ref{table:CO-dialog-prompt} in Appendix~\ref{meme_representation_generation_prompts_sec}). 

In this design, the model produces a structured conversation between two speakers: \textit{Host1}, who provides explanatory content, and \textit{Host2}, who offers reactions, questions, or informal interpretations. The dialog is intentionally concise, with meaning distributed across turns to support gradual understanding and anticipation, and concludes with a punchline-style summary that conveys the meme’s intent. The prompt also specifies TTS constraints (e.g., distinct voice styles and genders) to ensure clear speaker separation. Outputs are generated in a structured JSON format containing both dialog content and TTS specifications, enabling consistent downstream rendering.

\subsection{Commentator-Character (CH) Design}

To enhance immersion, participants proposed incorporating a character from the meme as an active speaker in the conversation. We operationalized this idea through the CH design, implemented via prompt constraints (Table~\ref{table:CH-dialog-prompt} in Appendix~\ref{meme_representation_generation_prompts_sec}).

In this configuration, the dialog occurs between a \textit{Host} (commentator) and a \textit{Guest} (a character from the meme). The Guest may adopt a first-person perspective or respond in a manner consistent with the character’s role, identity, or persona. The prompt enforces explicit self-introduction of the Guest to maintain role clarity, and guides the interaction such that the character meaningfully contributes to the narrative and punchline. Similar to CO, the dialog remains concise and structured, and includes TTS instructions specifying distinct and contextually appropriate voice characteristics. Compared to CO, which prioritizes interpretability through neutral commentary, CH emphasizes \emph{expressiveness and narrative immersion}. 

\subsection{Caption-style Design (Baseline)}

We also include a caption-style baseline (BL) in \sysname{} to reflect current practices and provide a reliable fallback. This design was strongly recommended by participants for scenarios requiring maximal clarity and explicitness.

The BL prompt (Table~\ref{table:BL-description-prompt} in Appendix~\ref{meme_representation_generation_prompts_sec}) generates a single-speaker narration delivered by a \textit{Narrator}. The narration follows a structured progression: identifying the meme template (when applicable), describing visual elements in detail (e.g., layout, expressions, embedded text), and concluding with an explicit explanation of the meme’s intent. The prompt encourages a natural, conversational tone while maintaining completeness and self-containment. It also specifies a consistent TTS configuration (warm, descriptive voice) to ensure intelligibility. In contrast to dialog-based designs, BL prioritizes \emph{semantic completeness} over interaction and interpretive engagement.

\subsection{Offline \sysname{} Evaluation}

Following standard practice in dialog and LLM-output evaluation, we conducted an offline human assessment of \sysname{}-generated meme representations. The evaluation was designed to address the following research question: \textbf{RQ1:} \textit{How well do the audio representations capture the intent, cultural nuances, and contextual details of a meme?}

\subsubsection{Dataset Collection and Annotation}

We constructed a dataset of $100$ meme images to support the offline evaluation, ensuring diversity in structure, content, and style. The dataset was compiled from two sources: the Kaggle Meme Images Dataset~\cite{kaggle2026memeimagesdataset} and the Imgflip platform~\cite{imgflip2026imgflip}. 

The Kaggle dataset, originally curated for sentiment analysis, contains a mixture of general images and memes. To ensure relevance, we filtered this dataset to retain only images that conform to common meme characteristics (e.g., presence of overlaid text, recognizable templates, or multimodal humor). From this filtered subset, we randomly sampled $62$ memes. To complement this, we additionally collected $38$ popular meme instances from Imgflip, prioritizing templates with high usage frequency and cultural familiarity.

We ensured that all selected memes contained both visual and textual components, as such multimodal interplay is central to meme interpretation and poses known challenges for blind users. We excluded memes containing hateful, abusive, or politically sensitive content to avoid confounds.

Each meme was manually annotated using a structured schema (Table~\ref{table:annotation_schema} in Appendix~\ref{meme_annotation_schema_sec}) designed to capture both low-level visual details and high-level semantic intent. Specifically, annotations included template identity, panel structure, key visual elements, verbatim textual content, and a concise intent summary describing the humor mechanism. Template names were validated against popular meme repositories (e.g., Imgflip, Kapwing) to ensure consistency and correctness.

To improve annotation reliability, two annotators independently labeled each meme, followed by a reconciliation process to resolve disagreements through discussion. This process ensured consistency in interpreting subjective fields, such as \texttt{intent summary}, which are critical for evaluating generated outputs. The resulting dataset provides a grounded reference for assessing the quality and fidelity of \sysname{} representations.

\subsubsection{Models and Metrics}

We evaluated \sysname{} representations generated using multiple LLMs (with temperature set to 0 to minimize random results). Specifically, we experimented with: (i) Gemini 3 Flash, (ii) Gemini 3 Pro, (iii) GPT-5.1, and (iv) Gemma 4. For each meme, all models were prompted using the same instructions (Tables~\ref{table:BL-description-prompt}, \ref{table:CO-dialog-prompt}, and \ref{table:CH-dialog-prompt}) for each representation type (BL, CO, CH) to ensure consistency.

Evaluating dialog-like and multimodal outputs remains challenging, as standard n-gram metrics (e.g., BLEU) fail to capture semantic and experiential quality~\cite{lowe2016evaluation}. Recent work, particularly the FED framework, has advocated for fine-grained human evaluation along multiple dimensions, such as coherence, engagement, grounding, and correctness~\cite{mehri2020unsupervised}. Guided by this framework, we define the following metrics tailored to meme understanding: (i) \textbf{\textit{Intent Clarity}}: How easily a listener can understand the meme’s setup, punchline, and tone; represented as a 5-point Likert item (1 = very difficult, 5 = very easy); (ii) \textbf{\textit{Representation Fidelity}}: Whether humor-critical visual elements (e.g., expressions, relationships, scene structure) are correctly captured; binary representation (0 = incorrect/missing, 1 = correct); (iii) \textbf{\textit{Text Accuracy}}: Whether on-image text is reproduced or paraphrased accurately; binary representation (0 = incorrect, 1 = correct); and (iv) \textbf{\textit{Speculation}}: Presence of unsupported, irrelevant, or hallucinated content; binary representation (0 = absent, 1 = present).

Each <meme, model, representation> triplet was independently evaluated by two annotators who were not part of the focus group, improving external validity. The annotators were provided with the original meme, generated outputs, and the dataset annotations (Table~\ref{table:annotation_schema}) as grounding reference. Items were presented in randomized order via a custom web interface to mitigate ordering effects.

We computed inter-annotator agreement for each metric (Cohen’s $\kappa$ for binary measures and weighted $\kappa$ for Likert ratings). Rather than resolving disagreements, we retained annotator variability by aggregating ratings across annotators, following prior work~\cite{mehri2020unsupervised, deriu2021survey}. Specifically, Likert ratings were averaged and binary metrics were reported as proportions. All subsequent analyses were conducted on these aggregated scores.

\subsubsection{Results}

\begin{table*}[t!]
\centering
{\small
{\def\arraystretch{1.1}
\begin{tabular}{lcccccccccccc}
  \toprule
  \multirow{2}{*}{\textbf{Model}} & \multicolumn{3}{c}{\textbf{Intent Clarity} $\uparrow$} & \multicolumn{3}{c}{\textbf{Representation Fidelity} $\uparrow$} & \multicolumn{3}{c}{\textbf{Text Accuracy} $\uparrow$} & \multicolumn{3}{c}{\textbf{Speculation} $\downarrow$} \\
  \cmidrule(lr){2-4} \cmidrule(lr){5-7} \cmidrule(lr){8-10} \cmidrule(lr){11-13}
  & \textbf{BL} & \textbf{CO} & \textbf{CH} & \textbf{BL} & \textbf{CO} & \textbf{CH} & \textbf{BL} & \textbf{CO} & \textbf{CH} & \textbf{BL} & \textbf{CO} & \textbf{CH} \\
  \midrule
  Gemini 3 Flash & \textbf{4.20} & 4.12 & 3.94 & \textbf{0.91} & 0.89 & 0.88 & \textbf{0.97} & 0.96 & 0.94 & \textbf{0.07} & 0.08 & 0.09 \\
  Gemini 3 Pro & \textbf{4.51} & 4.42 & 4.27 & \textbf{0.93} & \textbf{0.93} & 0.91 & \textbf{0.98} & 0.97 & 0.95 & \textbf{0.05} & 0.07 & 0.07 \\
  GPT-5.1 & \textbf{4.43} & 4.32 & 4.10 & \textbf{0.95} & 0.94 & 0.91 & \textbf{0.99} & 0.96 & 0.93 & \textbf{0.04} & 0.08 & 0.10 \\
  Gemma 4 & \textbf{3.68} & 3.46 & 3.23 & \textbf{0.82} & 0.77 & 0.80 & \textbf{0.94} & 0.91 & 0.88 & \textbf{0.12} & 0.15 & 0.19 \\
  \bottomrule
\end{tabular}
}
}
\vspace{-7pt}
\caption{Offline evaluation results aggregated across 100 memes and two annotators. Intent Clarity is a mean Likert score; Representation Fidelity, Text Accuracy, and Speculation are proportions (higher is better except Speculation).}
\label{tab:offline_results}
\end{table*}

Table~\ref{tab:offline_results} summarizes results across 100 memes and two annotators. Inter-annotator agreement was strong (Intent Clarity: $\kappa=0.78$; Representation Fidelity: $0.71$; Text Accuracy: $0.86$; Speculation: $0.82$), supporting reliability (variance statistics in Appendix~\ref{offline_evaluation_details_sec}).
Gemini Pro and GPT performed the best, followed by Gemini Flash, with Gemma trailing. BL and CO exhibited high \textit{Intent Clarity}, while CH was lower across all models. In contrast, \textit{Fidelity} and \textit{Text Accuracy} remained high across all conditions.

Friedman tests showed a significant effect of condition on \textit{Intent Clarity} for all models (Gemini Flash: $\chi^2(2)=28.34$, $p<.001$, $W=0.142$; Gemini Pro: $\chi^2(2)=9.64$, $p=.008$, $W=0.048$; GPT: $\chi^2(2)=19.18$, $p<.001$, $W=0.096$; Gemma: $\chi^2(2)=29.49$, $p<.001$, $W=0.147$), with small effect sizes. Post-hoc Wilcoxon signed-rank tests (with Holm's correction) showed CH was significantly worse than BL and CO for Gemini Pro, Gemini Flash, and GPT; for Gemma, all pairwise differences were significant (BL $>$ CO $>$ CH). No significant effects were observed for \textit{Fidelity} and \textit{Speculation}; for \textit{Text Accuracy}, marginal omnibus effects were observed for GPT and Gemma, but these did not yield significant pairwise differences.

Absolute differences in \textit{Intent Clarity} were small (typically $\leq 0.3$), indicating limited practical degradation. CO closely matched BL, especially for stronger models, showing that dialog structuring preserves semantic quality when coherence is maintained. In contrast, CH exhibited lower clarity and higher speculation, suggesting added complexity from character role modeling and turn alignment.

Overall, results indicate that dialog representations largely preserve comprehensibility, with modest degradation mainly in CH. We next examine if these structural differences impact \emph{engagement and enjoyment} via a user study with blind users.
\section{User Study with Blind Users}

We conducted an IRB-approved user study with blind participants to evaluate \sysname{} and address the following research questions: \textbf{RQ2:} \textit{What are the strengths and limitations of dialog-style meme representations compared to caption-style descriptions?} and \textbf{RQ3:} \textit{How can dialog-style representations be improved to enhance engagement and enjoyment?}

\subsection{Study Design}

We recruited 14 blind participants through email lists and word-of-mouth. All participants were familiar with AI-based image description tools and reported regular exposure to memes. Table~\ref{table:participants} (Appendix~\ref{userstudy_participant_demographics_sec}) shows the demographics of participants.

We used a within-subject design with three conditions: 

\begin{itemize}[noitemsep]
  \item \textbf{Baseline (BL):} caption-style audio description by a single speaker.
  \item \textbf{Commentator--Commentator (CO):} dialog between two speakers. 
  \item \textbf{Commentator--Character (CH):} dialog between a speaker and a meme character.
\end{itemize}

The BL condition reflects the dominant paradigm used by current AI-based accessibility tools, where images are described through single-speaker, caption-style narration. As such, it provides an ecologically valid and deployment-relevant baseline for comparison. Each participant completed 6 tasks (2 per condition), given the session time limit of 2 hours. In each task, the participant listened to a meme’s audio representation and completed a questionnaire measuring comprehension, emotional response, immersion, entertainment, funniness, and willingness to engage further (Table~\ref{table:userstudy_questions}). Q1 was open-ended; Q2--Q6 used 5-point Likert scales adapted from prior work~\cite{chen2025memespeak,xu2024memory,obrien2018practical,richardson2018measuring, nabi2006emotional,ryan1983relation}. These measures capture dialog effectiveness from a user perspective, reflecting perceived coherence, pacing, and conversational engagement rather than purely structural dialog metrics. We used Gemini 3 Pro to generate the representations, given its best performance in the offline evaluation.

\begin{table}[b!]
\centering
{\small
{\def\arraystretch{1.1}
\begin{tabular}{ m{7.2cm} }
\toprule
  \multicolumn{1}{c}{\textbf{Questions}} \\
  \midrule
  \textbf{Q1:} What is the author trying to convey in the meme?\\
  \textbf{Q2:} Listening to this audio, my emotions frequently \newline fluctuated (such as feeling happy, sad, angry, etc.).\\
  \textbf{Q3:} I was deeply immersed in this audio narrative \newline experience.\\
  \textbf{Q4:} I found the meme audio entertaining.\\
  \textbf{Q5:} The audio format enhanced the funniness of the meme.\\
  \textbf{Q6:} I would like to listen to more memes like this.\\
  \bottomrule
\end{tabular}
}
}
\vspace{-6pt}
\caption{User study questions.}
\label{table:userstudy_questions}
\end{table}

To control for ordering and learning effects, we used six distinct memes (Appendix~\ref{memes_userstudy_sec}) and counterbalanced study conditions using a Latin square design~\cite{bradley1958complete}. The selected memes were self-contained and required only common background knowledge, minimizing the influence of participants' prior familiarity with specific topics and enabling a fair comparison across conditions. The study was conducted in person using a Windows laptop with JAWS and NVDA installed. Following a brief practice session, participants completed all six tasks and then participated in a semi-structured exit interview. Each session lasted approximately two hours, and participants received an \$80 gift card as compensation.

\subsection{Measures and Analysis}

With participants' written consent, we collected questionnaire responses, think-aloud observations, and interview data. For Q1, responses were manually coded for correctness against the intended meme meaning. For Q2--Q6 (ordinal repeated measures), we used Friedman tests followed by Wilcoxon signed-rank post-hoc tests with Holm correction.
Qualitative data were analyzed using open and axial coding~\cite{saldana2011fundamentals}. Two authors independently conducted initial coding, resolved discrepancies through discussion, and collaboratively derived higher-level themes.

\subsection{Results}

\begin{figure*}[!t]
\centering
  \includegraphics[trim={0.4cm 0.2cm 0.4cm 0.2cm},clip,width=0.95\textwidth]{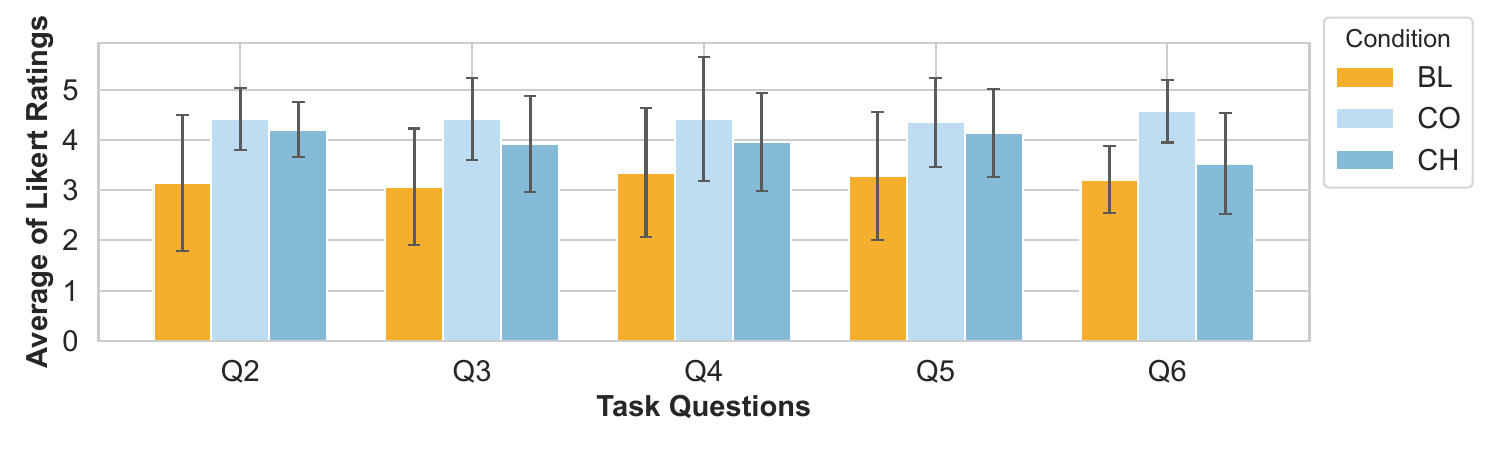}
  \vspace{-10pt}
  \caption{Average Likert ratings (Q2 -- Q6) in the questionnaire. All effects significant ($p<.001$). CO consistently outperforms BL, and CH shows intermediate gains.}
  \label{fig:likert-barchart}
\end{figure*}

\noindent\textbf{Comprehension (Q1).}
We coded open-ended responses against the intended meme meaning (binary correct/incorrect). Two annotators achieved high agreement (Cohen’s $\kappa=0.87$), with disagreements resolved by a third annotator.

All conditions achieved high comprehension: BL (92.85\%), CO (89.28\%), and CH (82.14\%). BL performed best, likely due to its explicit and coverage-oriented narration. CO remained close to BL, suggesting that distributing information across dialog turns does not substantially affect interpretability. CH showed a modest decrease, which may reflect increased ambiguity in speaker roles and turn coherence. Importantly, comprehension remained above 80\% across all conditions.

\vspace{0.4em}
\noindent\textbf{Engagement, Affect, and Preference (Q2--Q6).}
Results are shown in Figure~\ref{fig:likert-barchart}. We analyzed Likert responses using Friedman tests with Wilcoxon signed-rank post-hoc tests (Holm-corrected); full statistics are reported in Appendix~\ref{userstudy_full_inferential_statistical_results_sec}. Friedman tests indicated a significant effect of condition for all measures ($p<.001$). However, post-hoc comparisons revealed more nuanced patterns.

\textit{Key patterns:}
(i) \textbf{Emotional response (Q2)} increased significantly under CO and CH compared to BL; (ii) \textbf{Immersion (Q3)} improved under CO and CH, with CO outperforming BL and CH; (iii) \textbf{Entertainment and funniness (Q4--Q5)} showed selective improvements: CO significantly outperformed CH (Q4) and BL (Q5), while other pairwise differences were not consistently significant; (iv) \textbf{Preference (Q6)} showed the strongest effect, with CO significantly outperforming both BL and CH, while CH did not significantly differ from BL.

Overall, CO consistently yielded the highest ratings and most robust improvements across measures, whereas CH showed more variable gains. These results suggest that engagement benefits arise not only from multi-speaker dialog, but also from maintaining conversational coherence. Collectively, the findings indicate that structural re-representation plays a central role in shaping affective and experiential aspects of meme consumption.

\vspace{0.4em}
\noindent\textbf{Repeated Listening and Interaction Effort.}
Participants replayed memes under the CO condition more frequently for enjoyment, whereas replays under BL were primarily for clarification, based on think-aloud observations. Informally, CO appeared to reduce perceived effort by segmenting information into conversational turns. CH exhibited mixed patterns, with replays occurring both for enjoyment and for disambiguation when speaker roles were unclear. These patterns align with Q3--Q6 outcomes and suggest that \emph{turn-level segmentation} contributes to engagement.

\vspace{0.4em}
\noindent\textbf{Qualitative Insights.}
Six themes emerged from interview data analysis.

\noindent\textit{(i) Two speakers improve engagement.}\,
Participants consistently preferred dual-speaker dialog over single-speaker narration, citing reduced monotony, improved pacing, and clearer segmentation of information. Turn-taking helped distribute content into manageable units, making it easier to follow and less cognitively demanding. Participants also noted that dialog introduced anticipation, as they expected reactions or follow-ups from the second speaker. As one participant remarked, \textit{“It feels like when two people are talking, it’s not boring anymore.”} These benefits were most consistently observed in the CO condition, where conversational flow was clear and predictable; CH also improved engagement over BL, but less reliably.

\noindent\textit{(ii) Emotional expressiveness and coherence jointly shape experience.}\,
Participants emphasized that engagement depends not only on having multiple voices, but also on how expressively and coherently those voices interact. While CH introduced character-driven dialog, its effectiveness was sometimes limited by flat prosody or unclear conversational roles. Participants suggested richer emotional cues (e.g., sarcasm, emphasis, pacing) and more consistent voice behavior to make interactions feel natural. For example, one participant noted, \textit{“If the character sounded more expressive, it would feel like a real clip.”} These observations help explain why CH condition improved engagement over BL condition but did not match the consistency of CO condition, where speaker roles were clearer and interactions more coherent.

\noindent\textit{(iii) Dialog encourages an interactive mental model.}\,
Participants desired to engage more actively with dialog-style representations, such as replaying specific turns, asking follow-up questions, or probing unclear parts. This suggests that dialog naturally shifts users from passive consumption toward a more interactive interpretation process. As one participant put it, \textit{“I wish I could jump in and ask what they meant.”} This expectation of interaction was less evident in the BL condition, which was perceived as more static and explanatory.

\noindent\textit{(iv) Applicability beyond memes.}\,
Participants noted that dialog-style representations could extend beyond memes to other forms of visual content, particularly personal photos. They felt that dialog could better convey emotional context and lived experience compared to static descriptions. One participant explained, \textit{“This would be great for my photos ... it would bring back memories better than just descriptions.”} This suggests broader applicability of dialog-based representations for experiential accessibility.

\noindent\textit{(v) Avoid over-explanation.}\,
Participants preferred representations that preserve some level of inference rather than fully explaining the meme. BL descriptions were often perceived as overly explicit, reducing enjoyment by eliminating the need to interpret the humor. As one participant noted, \textit{“It tells me everything, so there’s nothing left to figure out.”} In contrast, dialog, particularly in CO, allowed meaning to unfold across turns, supporting humor through timing and partial revelation.

\noindent\textit{(vi) Need for customization.}\,
Participants highlighted the importance of flexibility in representation. They expressed interest in switching between BL, CO, and CH depending on context, as well as customizing voice styles and navigating content at finer granularity (e.g., replaying individual turns). One participant suggested, \textit{“I’d like to go back to just one part if I miss something.”} These preferences indicate that no single representation is universally optimal, and that adaptable interfaces may better support diverse user needs.
\section{Discussion}

Our results show that dialog-style representations can improve the non-visual experience of memes for blind users, while maintaining high comprehension (Q1). However, these improvements are not uniform across conditions or metrics, with the commentator--commentator (CO) design consistently yielding the most robust gains.

\vspace{0.4em}
\noindent\textbf{RQ2: Strengths and limitations.}
Dialog-style representations shift meme access from an \emph{informational} toward a more \emph{experiential} paradigm. While caption-style descriptions (BL) ensure semantic clarity~\cite{penuela2024investigating,gleason2019making}, they often fail to capture humor, timing, and cultural nuance. In contrast, dialog representation distributes meaning across turns, enabling users to infer intent through temporal unfolding. This is reflected in significant improvements in emotional response (Q2), immersion (Q3), and user preference (Q6), with more selective gains for entertainment and funniness (Q4--Q5).

These findings align with prior work on narrative engagement and experiential design, which emphasize temporal structure and anticipation~\cite{ryan1983relation,ghazarian2021discol}. Dialog introduces rhythm and contrast, allowing humor to emerge through interaction rather than explicit explanation.

At the same time, effectiveness depends critically on conversational coherence. The consistent advantage of CO over CH suggests that simply introducing additional expressive elements (e.g., meme characters) is insufficient. In CH, role ambiguity and breakdowns in turn coherence occasionally increased cognitive effort and reduced immersion. This highlights an important design principle: \emph{engagement arises not from expressiveness alone, but from structured and coherent interaction}; expressive elements that disrupt role clarity can reduce both interpretability and immersion.

\vspace{0.4em}
\noindent\textbf{RQ3: Improving engagement.}
Our findings suggest three directions for improvement.
First, \emph{emotional expressiveness} remains critical. Participants noted that richer prosody (e.g., sarcasm, emphasis) would improve realism and engagement. While prior work demonstrates benefits of emotional narration~\cite{chen2025memespeak}, our results suggest these gains may be amplified when applied at the level of dialog.
Second, dialog representations naturally invite \emph{interaction}. Participants expressed interest in asking follow-up questions and revisiting specific turns, suggesting that dialog-based systems should evolve toward interactive conversational interfaces. This aligns with emerging reliance on conversational visual assistants~\cite{pal2026visionaidqa, sindhu2025voice}.
Third, \emph{customization} is essential. Participants requested flexibility in representation style, voice characteristics, and navigation granularity. This indicates that future systems should support adaptive, user-controlled configurations rather than a single fixed representation.

\vspace{0.4em}
\noindent\textbf{Implications for Accessible Content Design.}
Our work extends prior research on accessible image descriptions~\cite{chen2025memespeak,xu2024memory}, which has largely focused on semantic comprehension. We show that \emph{structural re-representation}, specifically dialog-style rendering, plays a central role in enhancing engagement.

This perspective aligns with prior work emphasizing that user experience depends on both content and presentation~\cite{chen2025memespeak}. By modeling memes as conversations, \sysname{} better reflects their inherently social nature. More broadly, our findings suggest a shift from \emph{equivalence} (same information) to \emph{experiential parity} (comparable experience), particularly for socially and culturally rich media such as memes.

\vspace{0.4em}
\noindent\textbf{Limitations.}
Our work has several limitations. First, static voice configurations likely constrained the effectiveness of CH; future work should explore context-aware expressive speech. Second, we only considered English memes, limiting generalizability. Third, we focus on static images, excluding GIFs and videos where dialog representations may be even more effective. Fourth, each study session was limited to six meme tasks to fit within the two-hour session duration. Future longitudinal in-the-wild studies should evaluate dialog-style representations using substantially larger and more diverse meme collections spanning a broader range of topics, formats, and cultural contexts. Fifth, our user study sample size is modest, however, it is consistent with prior user studies involving blind participants, where recruitment is challenging and within-subject designs provide sufficient sensitivity for comparative evaluation. Sixth, we did not isolate the individual contributions of \sysname{}'s two key design elements, i.e., multiple voices and temporal unfolding, to user engagement. Although participants consistently identified both as important during the exit interviews, future work should disentangle their effects through controlled experiments to quantify their respective contributions to engagement, immersion, and enjoyment. Lastly, we note that contextual interpretation (e.g., cultural references) is inferred implicitly by the LLM and not externally grounded, which may introduce errors for niche memes.
\section{Conclusion}

We presented \sysname{}, a system that rethinks meme accessibility through dialog-style audio representations that model memes as structured two-speaker conversations. Across an offline evaluation and a user study with blind participants, we show that dialog-based representations maintain high semantic fidelity while improving engagement, immersion, and perceived humor over caption-style descriptions. These findings highlight a key limitation of current approaches: optimizing for comprehension alone is insufficient for socially and culturally rich media such as memes. Instead, conversational structure and temporal unfolding play a central role in shaping user experience. By moving toward \emph{experiential accessibility}, \sysname{} points to a promising direction for assistive systems that support both understanding and engagement.

\section*{Acknowledgments}
We thank anonymous reviewers for their insightful feedback. We also thank the participants and Dr. William Seiple for their support with the user study.


\bibliography{references.bib}

\newpage
\nobalance
\appendix


\clearpage
\onecolumn
\section{Meme Representation Generation Prompts} \label{meme_representation_generation_prompts_sec}
\begin{table*}[h!]
\centering
{\small
  \begin{tcolorbox}[colback=tcolorbox-background, colframe=tcolorbox-frame, title=\texttt{Caption-style Description (Baseline) Generation Prompt}]

\begin{Verbatim}[breaklines=true, breakanywhere=true, breaksymbol=]
You are tasked with explaining memes to a blind person in an engaging monologue as a single narrator, and also generating TTS voice style instructions.

## Rules for Output
- Output strictly in JSON with two top-level keys:
  - "dialogues" → array of objects with "speaker": "Narrator" and "text".
  - "tts_instructions" → array with one object describing how the narrator’s voice should sound.
- No extra narration, no text outside the JSON.

## Rules for Dialogue
- Always use "speaker": "Narrator".
- Start by recognizing or naming the meme template (if known).
- Describe what’s happening visually so a blind listener can clearly imagine it (panels, labels, expressions, text placement).
- Keep the tone casual, cohesive, and natural — like a friend explaining something funny. Add light filler words or small reactions so it doesn’t sound robotic.
- End with a short summary punchline that explains the joke in plain language.
- Keep total length around 25-75 seconds depending on meme complexity. Don’t pad unnecessarily.

## Rules for TTS Instructions
- Provide "tts_instructions" with one object:
  - "speaker": "Narrator"
  - "style": Warm, conversational, descriptive voice. Clear pacing, friendly tone, with a hint of humor.

---

# Example Input → Output

Input: Meme: Distracted Boyfriend with labels “me,” “responsibilities,” and “new hobby.”

Output:
{
  "dialogues": [
    {"speaker": "Narrator", "text": "Oh hey, this one’s the distracted boyfriend meme."},
    {"speaker": "Narrator", "text": "You’ve got the guy walking with his girlfriend, but his head’s turned to check out another woman passing by."},
    {"speaker": "Narrator", "text": "In this version, the girlfriend is labeled ‘responsibilities,’ the guy is ‘me,’ and the other woman is ‘new hobby.’"},
    {"speaker": "Narrator", "text": "So yeah, the joke is basically about ignoring your duties just because something fun comes along."}
  ],
  "tts_instructions": [
    {"speaker": "Narrator", "style": "Warm, conversational, descriptive voice. Clear pacing, friendly tone, with a hint of humor."}
  ]
}
\end{Verbatim}
  \end{tcolorbox}
}
\vspace{-10pt}
\caption{A prompt for generating caption-style description.}
\label{table:BL-description-prompt}
\end{table*}

\clearpage

\begin{table*}[h!]
\centering
{\small
  \begin{tcolorbox}[colback=tcolorbox-background, colframe=tcolorbox-frame, title=\texttt{Commentator-Commentator (CO) Dialog Generation Prompt}]

\begin{Verbatim}[breaklines=true, breakanywhere=true, breaksymbol=]
Your task is to describe a meme to a blind person in an engaging dialog format between two commentators, and also generate text-to-speech (TTS) voice style instructions. While creating the dialog and TTS instructions, you should strictly follow these rules:

Rules for constructing the dialog:

- The two commentators must be labeled "Host1" and "Host2".
- Host1 should start the conversation.
- Host1’s utterances should be more explanatory and Host2’s utterances should be more casual/reactive.
- The conversation should not be overly detailed. There should be room for self-interpretation for the listener.
- The conversation should be natural and concise: 25-75 seconds if narrated with a TTS. Don't pad content unnecessarily.
- The conversation should end with a plain punchline-style summary from Host2 to convey the meme intent such as humor or sarcasm.  

Rules for generating TTS instructions:

- Host1’s voice should have a clear, explanatory style.
- Host2’s voice should be more casual and reactive, complementing Host1’s voice. 
- Host 1 and Host 2 should not have the same gender.
- The voices of Host1 and Host2 should be distinct so the listener can easily tell them apart.
- Gender assignment should be either "male" or "female".

Rules for Output format:

- Output must be strictly a JSON object with two keys: "dialogs" and "tts_instructions".
- "dialogs" must be an array of objects, with each object having “speaker” and “text” fields along with their values.
- "tts_instructions" must be an array of objects, with each object having “speaker”, “style”, and “gender” fields along with their values.
- An illustrative example showing an input-output pair is given next. 

Example Input-Output pair:

Input: [Drake "Yes/No" meme image with labels "Working overtime" and "Taking a nap."]

Output:
{
  "dialogs": [
    {"speaker": "Host1", "text": "This is the classic Drake meme—two stacked panels showing preference."},
    {"speaker": "Host2", "text": "Oh yeah, the top panel has Drake waving his hand dismissively at 'working overtime.'"},
    {"speaker": "Host1", "text": "And the bottom shows Drake pointing approvingly at 'taking a nap.'"},
    {"speaker": "Host2", "text": "So basically: work bad, sleep good. The eternal struggle!"}
  ],
  "tts_instructions": [
    {"speaker": "Host1", "style": "Clear, explanatory tone. Professional but friendly, like a radio host.", "gender": "female"},
    {"speaker": "Host2", "style": "Casual, laid-back voice with good energy. More conversational and reactive.", "gender": "male"}
  ]
}
\end{Verbatim}
  \end{tcolorbox}
}
\vspace{-10pt}
\caption{A prompt for generating Commentator-Commentator (CO) dialog.}
\label{table:CO-dialog-prompt}
\end{table*}
\begin{table*}[h!]
\centering
{\small
  \begin{tcolorbox}[colback=tcolorbox-background, colframe=tcolorbox-frame, title=\texttt{Commentator-Character (CH) Dialog Generation Prompt}, boxsep=2pt, left=1pt, right=1pt, top=1pt, bottom=1pt]

\begin{Verbatim}[breaklines=true, breakanywhere=true, breaksymbol=]
Your task is to describe a meme to a blind person in an engaging dialog format between a commentator (i.e., Host) and a character (i.e., Guest) from the meme, and also generate text-to-speech (TTS) voice style instructions. While creating the dialog and TTS instructions, you should strictly follow these rules:

Rules for Dialog:

- The speakers must be labeled "Host" and "Guest".
- The Host should start the conversation.
- For Guest, if there are more than one characters in the meme, select the character that is associated with the punch-line that conveys intent like humor or sarcasm.
- The Guest must introduce themselves in their first utterance in the dialog. If the selected character/person from the meme has a known name, use it (e.g., "Hi, I'm Drake"). If the name is unknown, have the Guest describe themselves by relating to how the Host refers to them (e.g., if Host says "the person in the meme," Guest says "Yeah, I'm the person in the meme"). This helps blind listeners distinguish between speakers.
- If the meme suits first-person narration (e.g., character/person style meme), then the Guest should speak as if they are the character. If not, the guest should speak like a commentator.
- The conversation should not be overly detailed. There should be room for self-interpretation for the listener.
- The conversation should be natural and concise: 25-75 seconds if narrated with a TTS. Don't pad content unnecessarily.
- The conversation should end with a plain punchline-style summary conveying the meme intent such as humor or sarcasm. 

Rules for TTS Instructions:

- First decide the Guest's gender based on the meme character or context. If the gender of the character is not clear (e.g., a bird), choose the gender that creates better contrast between the two speakers.
- The gender of the Host should be the opposite of Guest to maximize voice distinction.
- The gender should be either “male” or “female”.
- The voice style for the Host should be warm and clear with an explanatory tone. The pacing should be moderate like a professional podcast narrator.
- The voice style for the Guest will depend on the meme character. If the selected meme character for the Guest is well known (e.g., a celebrity), then configure the style to match their publicly-known speech attributes (tone, personality, pacing). If the selected character is not a public figure, then use a friendly and conversational style that best complements the Host’s style.
- Try your best to make the voice profiles of Host and Guest distinct so the listener can easily tell them apart.

Rules for Output format:

- Output must be strictly a JSON object with two keys: "dialogs" and "tts_instructions".
- "dialogs" must be an array of objects, with each object having “speaker” and “text” fields along with their values.
- "tts_instructions" must be an array of objects, with each object having “speaker”, “style”, and “gender” fields along with their values.
- An illustrative example showing an input-output pair is given next. 

Example Input-Output pair:

Input: [Meme: Drake “Yes/No” with labels “Working overtime” and “Taking a nap.”]

Output: {
  "dialogs": [
    {"speaker": "Host", "text": "This is the Drake meme—the two stacked panels."},
    {"speaker": "Guest", "text": "Yeah, in the top I’m waving my hand like, ‘nope,’ at the words ‘working overtime.’"},
    {"speaker": "Host", "text": "And then in the bottom, you’re pointing happily at ‘taking a nap.’"},
    {"speaker": "Guest", "text": "Exactly—work is out, sleep is in."}
  ],
  "tts_instructions": [
    {"speaker": "Host", "style": "Warm, clear, explanatory tone. Moderate pace, like a friendly podcast narrator.", "gender": "female"},
    {"speaker": "Guest", "style": "Casual, smug, playful voice—try to sound like Drake, with a relaxed pace and confident energy.", "gender": "male"}
  ]
}
\end{Verbatim}
  \end{tcolorbox}
}
\vspace{-10pt}
\caption{A prompt for generating Commentator-Character (CH) dialog.}
\label{table:CH-dialog-prompt}
\end{table*}

\clearpage
\onecolumn
\section{Meme Annotation Schema} \label{meme_annotation_schema_sec}
\begin{table}[h]
\centering
{\small
{\def\arraystretch{1.1}
\begin{tabular}{ >{\centering\arraybackslash}m{1.8cm} >{\raggedright\arraybackslash}m{7.0cm}}
  \toprule
  \textbf{Field} & \textbf{Description}\\
  \midrule
  \texttt{template name} & Meme template identity (``unknown'' if a template is unrecognizable) \\ \arrayrulecolor{tableline-gray}\hline
  \texttt{panels} & Panel count and layout order \\ \arrayrulecolor{tableline-gray}\hline
  \texttt{key visuals} & Per-panel visual elements (e.g., characters, poses, expressions, objects) \\ \arrayrulecolor{tableline-gray}\hline
  \texttt{labels} & Exacted on-image text (case-sensitive, verbatim) \\ \arrayrulecolor{tableline-gray}\hline
  \texttt{intent summary} & Explanation of humor mechanism in a meme in 1–2 sentences \\
  \arrayrulecolor{black}
  \bottomrule
\end{tabular}
}
}
\vspace{-5pt}
\caption{Meme image annotation schema.}
\label{table:annotation_schema}
\end{table}

\onecolumn
\section{Additional Offline Evaluation Details}
\label{offline_evaluation_details_sec}

\subsection{Variance Statistics}

Table~\ref{tab:offline_variance} reports the standard deviations (SD) for \textit{Intent Clarity} and binomial standard errors (SE) for the binary metrics, computed across 100 memes and two annotators.
Variance is moderate overall, with slightly higher variability in CH across models, indicating less consistent performance for character-based dialog generation.

\begin{table*}[h]
\centering
{\small
{\def\arraystretch{1.1}
\begin{tabular}{ l l c c c c c c c c c c c c }
  \toprule
  \multirow{2}{*}{\textbf{Model}} & \multicolumn{3}{c}{\textbf{Intent Clarity (SD)}} & \multicolumn{3}{c}{\textbf{Representation Fidelity (SE)}} & \multicolumn{3}{c}{\textbf{Text Accuracy (SE)}} & \multicolumn{3}{c}{\textbf{Speculation (SE)}} \\
  \cmidrule(lr){2-4} \cmidrule(lr){5-7} \cmidrule(lr){8-10} \cmidrule(lr){11-13}
  & \textbf{BL} & \textbf{CO} & \textbf{CH} & \textbf{BL} & \textbf{CO} & \textbf{CH} & \textbf{BL} & \textbf{CO} & \textbf{CH} & \textbf{BL} & \textbf{CO} & \textbf{CH} \\
  \midrule
  Gemini 3 Flash & 0.61 & 0.66 & 0.72 & 0.029 & 0.031 & 0.032 & 0.017 & 0.020 & 0.024 & 0.026 & 0.028 & 0.029 \\
  Gemini 3 Pro & 0.52 & 0.58 & 0.64 & 0.026 & 0.026 & 0.029 & 0.014 & 0.017 & 0.022 & 0.022 & 0.026 & 0.026 \\
  GPT-5.1 & 0.54 & 0.60 & 0.68 & 0.022 & 0.024 & 0.029 & 0.010 & 0.020 & 0.025 & 0.020 & 0.028 & 0.030 \\
  Gemma 4 & 0.78 & 0.82 & 0.89 & 0.038 & 0.042 & 0.040 & 0.024 & 0.029 & 0.033 & 0.032 & 0.035 & 0.039 \\
  \bottomrule
\end{tabular}
}
}
\vspace{-5pt}
\caption{Variance statistics for offline evaluation metrics. Standard deviations (SD) are reported for \textit{Intent Clarity}; binomial standard errors (SE) are reported for binary metrics (\textit{Representation Fidelity}, \textit{Text Accuracy}, \textit{Speculation}).}
\label{tab:offline_variance}
\end{table*}

\subsection{Effect Sizes}

Table~\ref{tab:offline_effectsizes} shows Kendall’s $W$ for Friedman tests on \textit{Intent Clarity}.

\begin{table}[h]
\centering
{\small
{\def\arraystretch{1.1}
\begin{tabular}{ l c }
  \toprule
  \textbf{Model} & \textbf{Kendall’s $W$} \\
  \midrule
  Gemini 3 Flash & 0.142 \\
  Gemini 3 Pro & 0.048 \\
  GPT-5.1 & 0.096 \\
  Gemma 4 & 0.147 \\
  \bottomrule
\end{tabular}
}
}
\vspace{-5pt}
\caption{Effect sizes for condition differences in \textit{Intent Clarity}.}
\label{tab:offline_effectsizes}
\end{table}

All effect sizes are \textbf{small} ($W \leq 0.15$), indicating that although condition differences are statistically significant, their practical impact on comprehensibility is limited.

\subsection{Pairwise Effect Sizes}

Table~\ref{tab:offline_pairwise} reports representative Wilcoxon signed-rank effect sizes ($r$).

\begin{table}[h]
\centering
{\small
{\def\arraystretch{1.1}
\begin{tabular}{ l c }
  \toprule
  \textbf{Comparison} & \textbf{Effect Size ($r$)} \\
  \midrule
  BL vs. CO & 0.10–0.18 (small) \\
  BL vs. CH & 0.25–0.38 (small–moderate) \\
  CO vs. CH & 0.20–0.32 (small–moderate) \\
  \bottomrule
\end{tabular}
}
}
\vspace{-5pt}
\caption{Representative pairwise effect sizes for \textit{Intent Clarity}.}
\label{tab:offline_pairwise}
\end{table}

Pairwise comparisons show larger (but still modest) effects involving CH, consistent with reduced clarity in character-driven dialog.

\clearpage
\onecolumn
\section{User Study Participant Demographics} \label{userstudy_participant_demographics_sec}

\begin{table*}[h!]
\centering
{\small
{\def\arraystretch{1.1}
\begin{tabular}{ m{0.2cm} m{0.8cm} m{1.5cm} m{1.9cm} m{1.8cm} m{1.9cm} >{\raggedright\arraybackslash}m{2.2cm} >{\raggedright\arraybackslash}m{2.4cm}}
\toprule
 \multirow{2}{*}{\textbf{ID}} & \textbf{Age /} & \multirow{2}[1]{*}{\shortstack[l]{\textbf{Visual}\\ \textbf{Condition}}} & \multirow{2}[1]{*}{\shortstack[l]{\textbf{Web Browsing}\\ \textbf{Experience}}} & \multirow{2}[1]{*}{\shortstack[l]{\textbf{Preferred}\\ \textbf{Screen Reader}}} & \multirow{2}[1]{*}{\shortstack[l]{\textbf{Social Forums}\\ \textbf{Experience}}} & \multirow{2}[1]{*}{\shortstack[l]{\textbf{Meme}\\ \textbf{Exposure}}} & \multirow{2}{*}{\textbf{AI Assistive Tools}}\\
 & \textbf{Gender} & & & & & & \\
 \midrule
 P1 & 44/M & Blind & Daily & JAWS, NVDA & 2-3 times/week & X, Reddit & Be My AI, ChatGPT\\ \arrayrulecolor{tableline-gray}\hline
 P2 & 32/M & Blind & Daily & JAWS & Daily & Yahoo, Reddit & ChatGPT\\ \arrayrulecolor{tableline-gray}\hline
 P3 & 25/F & Blind & Daily & JAWS, NVDA & 4-5 times/week & WhatsApp & ChatGPT\\ \arrayrulecolor{tableline-gray}\hline
 P4 & 56/F & Low Vision & 2-3 times/week & JAWS & 0-1 time/week & Facebook & JAWS Picture Smart AI\\ \arrayrulecolor{tableline-gray}\hline
 P5 & 66/M & Blind & 1-2 times/week & JAWS & 1 time/week & WhatsApp & Be My AI\\ \arrayrulecolor{tableline-gray}\hline
 P6 & 52/F & Low Vision & Daily & JAWS & 2-4 times/week & Messages, Reddit & ChatGPT\\ \arrayrulecolor{tableline-gray}\hline
 P7 & 41/F & Blind & Daily & JAWS & Daily & Facebook & Be My AI\\ \arrayrulecolor{tableline-gray}\hline
 P8 & 38/F & Blind & Daily & JAWS, VoiceOver & 4 times/week & X, Bluesky & Be My AI\\ \arrayrulecolor{tableline-gray}\hline
 P9 & 29/M & Light \newline Perception & Daily & JAWS, NVDA & Daily & Messages, WhatsApp & Be My AI, JAWS Picture Smart AI\\ \arrayrulecolor{tableline-gray}\hline
 P10 & 22/F & Blind & Daily & NVDA & Daily & Facebook, Yahoo & Be My AI, ChatGPT\\\arrayrulecolor{tableline-gray}\hline
 P11 & 46/M & Blind & 4-5 times/week & JAWS & 2-3 times/week & Facebook & Be My AI, JAWS Picture Smart AI\\ \arrayrulecolor{tableline-gray}\hline
 P12 & 30/M & Light \newline Perception & Daily & JAWS & Daily & Facebook, Reddit & ChatGPT\\ \arrayrulecolor{tableline-gray}\hline
 P13 & 23/F & Blind & Daily & NVDA & Daily & Messages, Facebook & Be My AI, ChatGPT\\ \arrayrulecolor{tableline-gray}\hline
 P14 & 27/M & Blind & Daily & JAWS & Daily & X, Yahoo & Be My AI\\ 
 \arrayrulecolor{black}
 \bottomrule
\end{tabular}
}
}
\vspace{-5pt}
\caption{Demographics of blind participants in the user study. All information was self-reported by the participants.}
\label{table:participants}
\end{table*}

\section{Memes in User Study Evaluation} \label{memes_userstudy_sec}
\begin{figure*}[h!]
\centering
  \includegraphics[trim={0.0cm 0.0cm 0.0cm 0.0cm},clip,width=0.95\textwidth]{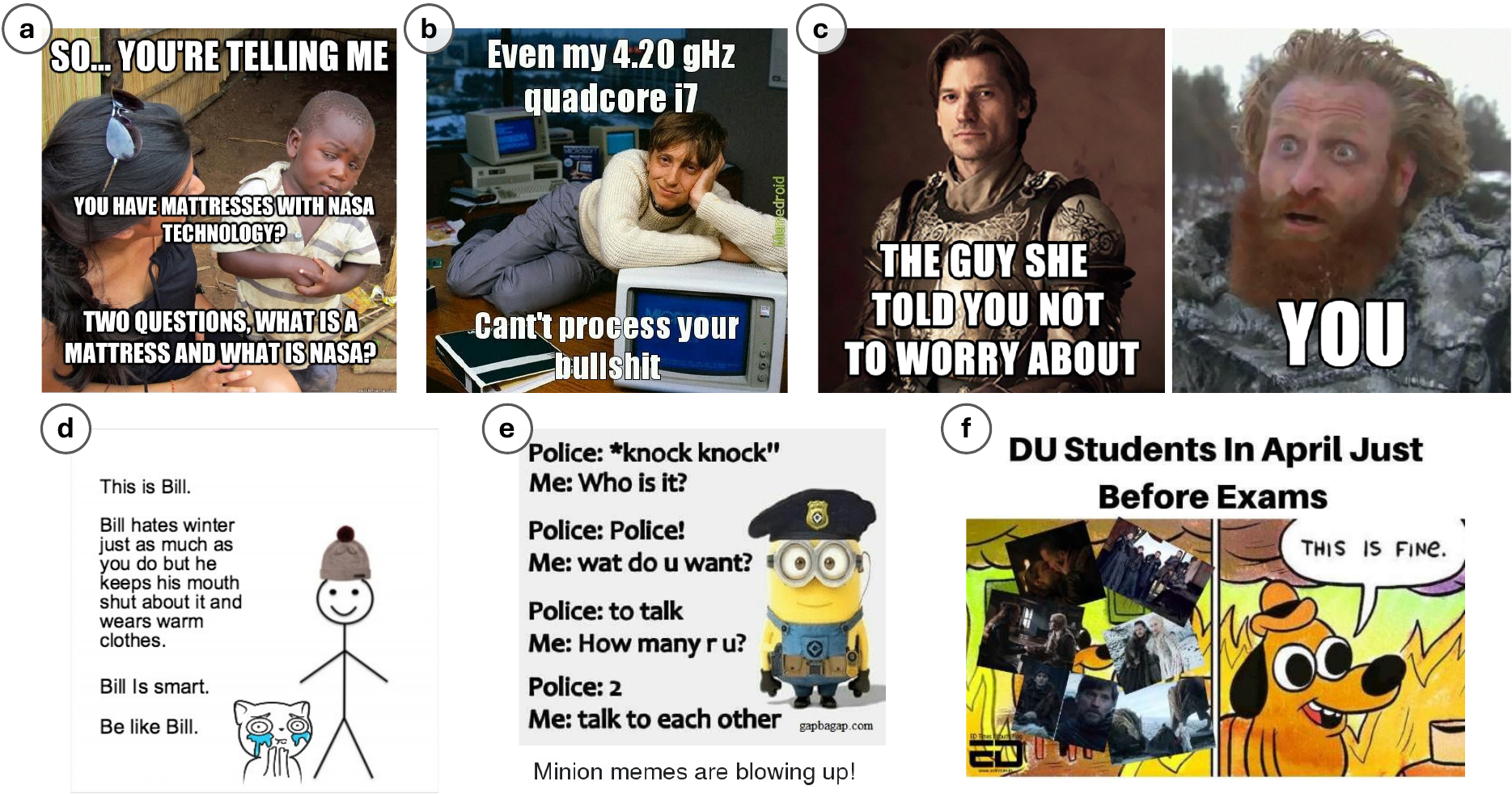}
  \vspace{-5pt}
  \caption{Image memes used in the user study.}
  \label{fig:userstudy_memes}
\end{figure*}

The six memes shown in Figure~\ref{fig:userstudy_memes} were used in the user study. 

\clearpage
\twocolumn
\section{Full Inferential Statistical Results (Q2--Q6)}\label{userstudy_full_inferential_statistical_results_sec}


We analyzed Likert-scale responses (Q2--Q6) using Friedman tests followed by pairwise Wilcoxon signed-rank tests with Holm correction for multiple comparisons. Effect sizes are reported as Kendall’s $W$ for omnibus tests and $r$ for pairwise comparisons.
The inferential statistical results for Q2--Q6 are shown in Table~\ref{table:inferential_statistical_results}.

\begin{table}[h!]
\centering
{\small
{\def\arraystretch{1.1}
\begin{tabular}{ c c c c }
  \toprule
  \rowcolor{light-gray}
  \multicolumn{4}{l}{\textbf{Q2: Emotional fluctuation}}\\
  \midrule
  \multicolumn{4}{l}{\textbf{Friedman test}}\\
  \multicolumn{4}{l}{$\chi^2(2)=21.23$, $p<.001$, $W=0.38$}\\
  \arrayrulecolor{black}\hline
  \multicolumn{4}{c}{\textbf{Post-hoc (Wilcoxon signed-rank test)}}\\
  \textbf{Comparison} & \textbf{$W$} & \textbf{$p$-value} & \textbf{Effect Size ($r$)} \\
  \midrule
  BL vs. CO & $0.0$ & $.0012$ & $.89$\\
  BL vs. CH & $13.5$ & $.0030$ & $.75$\\
  CO vs. CH & $3.5$ & $.0578$ & $.72$\\

  \toprule
  \rowcolor{light-gray}
  \multicolumn{4}{l}{\textbf{Q3: Immersion}}\\
  \midrule
  \multicolumn{4}{l}{\textbf{Friedman test}}\\
  \multicolumn{4}{l}{$\chi^2(2)=16.24$, $p<.001$, $W=0.29$}\\
  \arrayrulecolor{black}\hline
  \multicolumn{4}{c}{\textbf{Post-hoc (Wilcoxon signed-rank test)}}\\
  \textbf{Comparison} & \textbf{$W$} & \textbf{$p$-value} & \textbf{Effect Size ($r$)} \\
  \midrule
  BL vs. CO & $35.0$ & $.0027$ & $.68$\\
  BL vs. CH & $90.0$ & $.0414$ & $.43$\\
  CO vs. CH & $60.0$ & $.0414$ & $.49$\\

  \toprule
  \rowcolor{light-gray}
  \multicolumn{4}{l}{\textbf{Q4: Entertainment}}\\
  \midrule
  \multicolumn{4}{l}{\textbf{Friedman test}}\\
  \multicolumn{4}{l}{$\chi^2(2)=16.48$, $p<.001$, $W=0.29$}\\
  \arrayrulecolor{black}\hline
  \multicolumn{4}{c}{\textbf{Post-hoc (Wilcoxon signed-rank test)}}\\
  \textbf{Comparison} & \textbf{$W$} & \textbf{$p$-value} & \textbf{Effect Size ($r$)} \\
  \midrule
  BL vs. CO & $70.0$ & $.1268$ & $.40$\\
  BL vs. CH & $74.0$ & $.2311$ & $.27$\\
  CO vs. CH & $14.0$ & $.0140$ & $.73$\\

  \toprule
  \rowcolor{light-gray}
  \multicolumn{4}{l}{\textbf{Q5: Funniness}}\\
  \midrule
  \multicolumn{4}{l}{\textbf{Friedman test}}\\
  \multicolumn{4}{l}{$\chi^2(2)=18.00$, $p<.001$, $W=0.32$}\\
  \arrayrulecolor{black}\hline
  \multicolumn{4}{c}{\textbf{Post-hoc (Wilcoxon signed-rank test)}}\\
  \textbf{Comparison} & \textbf{$W$} & \textbf{$p$-value} & \textbf{Effect Size ($r$)} \\
  \midrule
  BL vs. CO & $64.0$ & $.0391$ & $.51$\\
  BL vs. CH & $61.0$ & $.0623$ & $.46$\\
  CO vs. CH & $9.0$ & $.0833$ & $.58$\\

  \toprule
  \rowcolor{light-gray}
  \multicolumn{4}{l}{\textbf{Q6: Willingness to consume more}}\\
  \midrule
  \multicolumn{4}{l}{\textbf{Friedman test}}\\
  \multicolumn{4}{l}{$\chi^2(2)=29.56$, $p<.001$, $W=0.53$}\\
  \arrayrulecolor{black}\hline
  \multicolumn{4}{c}{\textbf{Post-hoc (Wilcoxon signed-rank test)}}\\
  \textbf{Comparison} & \textbf{$W$} & \textbf{$p$-value} & \textbf{Effect Size ($r$)} \\
  \midrule
  BL vs. CO & $0.0$ & $<.001$ & $.90$\\
  BL vs. CH & $61.0$ & $.1578$ & $.32$\\
  CO vs. CH & $0.0$ & $<.001$ & $.90$\\
  \bottomrule
\end{tabular}
}
}
\vspace{-5pt}
\caption{Inferential statistical results for Q2--Q6.}
\label{table:inferential_statistical_results}
\end{table}

\end{document}